\documentclass[aps,prd,a4paper,eqsecnum,superscriptaddress,nofootinbib,showpacs,twocolumn,showkeys,amsfonts,amssymb,amsmath]{revtex4}

\usepackage{amssymb,latexsym}
\usepackage{amsmath, amsthm}
\usepackage{amscd}
\usepackage{amsthm}                
\usepackage{times}
\usepackage{epsfig}
\usepackage{psfrag}
\usepackage{graphicx}
\usepackage{amssymb,latexsym}

\numberwithin{equation}{section}

\begin{document}

\title{Gravitational collapse with tangential pressure}

\author{Daniele Malafarina} \email{daniele.malafarina@polimi.it}
\affiliation{Tata Institute of Fundamental Research, Homi Bhabha Road,
Colaba, Mumbai 400005, India}
\author{Pankaj S. Joshi} \email{psj@tifr.res.in}
\affiliation{Tata Institute of Fundamental Research, Homi Bhabha Road,
Colaba, Mumbai 400005, India}

\swapnumbers

\begin{abstract}
Using the general formalism for spherical gravitational
collapse developed in
\cite{PSJ},
we investigate here the final fate of a spherical distribution
of a matter cloud, where radial pressures vanish but tangential
pressures are non-zero.
Within this framework, firstly we examine the effect of
introducing a generic small pressure in a well-known black hole
formation process, which is that of an otherwise pressure-free
dust cloud. The intriguing result we find is that a dust collapse
that was going to a black hole final state could now go to
a naked singularity final configuration, when arbitrarily small
tangential pressures are introduced. The implications of such
a scenario are discussed in some detail. Secondly, the approach here
allows us to generalize the earlier results obtained on
gravitational collapse with non-zero tangential pressure, in the
presence of a non-zero cosmological constant. Finally, we discuss
the genericity of black hole and naked singularity formation
in collapse with non-zero tangential pressure. The treatment here
gives a unified and complete picture on collapse final states,
in terms of black hole and naked singularity formation, generalizing
the earlier results obtained for this class of collapse models.
Thus the role of tangential stresses towards determining collapse
endstates emerges in a straightforward and transparent
manner in our treatment.

\end{abstract}


\maketitle

\section{Introduction}

Since the first proposal of the Cosmic Censorship Conjecture (CCC)
\cite{Penrose},
a great effort has been devoted in understanding the mechanism
by which black holes and naked singularities can form as the end state
of the complete gravitational collapse of a massive object. However,
no general proof, or even a suitable mathematically rigorous formulation
of the CCC is available despite the efforts of many decades. Therefore
it has become essential and inevitable that gravitational collapse
within the framework of general relativity be studied in detail.

The current situation is that, despite the great amount of efforts
devoted in the past years in studying the CCC and its implications, the
issue of final fate of complete gravitational collapse of a massive body
is far from being entirely resolved. It is clear now how naked singularities
have to be considered as a general feature of general relativistic physics,
and that they may appear as the end-state of collapse in a broad
variety of situations (see e.g.
\cite{Ref}
and references therein).
Still, it is not yet fully understood how different kinds of matter lead to
the formation of either a black hole or a naked singularity. To give an example,
how the presence of pressure in the form of tangential stresses and
radial stresses might direct
the collapse towards the formation of black holes or naked
singularities is still a matter of debate.

The first gravitational collapse scenario studied in detail
within the framework of the general theory of relativity was the so called
Oppenheimer-Snyder-Datt model
\cite{OSD},
that represents a collapsing ball made of homogeneous pressureless
dust. It is seen here that an event horizon forms
before the appearance of the final spacetime singularity at the end of collapse,
therefore hiding the singularity inside a black hole. Further studies of
inhomogeneous dust collapse though, such as those in the
Lemaitre-Tolman-Bondi (LTB) models
\cite{LTB},
showed that naked singularities can indeed form during collapse, thus
violating the cosmic censorship
\cite{dust}.
However, the dust models are idealized scenarios
that ignore pressures within the cloud.
Therefore, collapse models with pressures, which are more
physically realistic scenarios, were examined and it was soon discovered
that for the collapse with non-zero pressures also, the end
state is just as likely to be a naked singularity, as for models
without pressures (see e.g.
\cite{fluid}).

In particular, collapse models with a vanishing radial
pressure, but with non-vanishing tangential stresses, have been
studied in some detail
\cite{press}.
These models are of particular interest, because they provide
a somewhat clear insight into the role of pressure towards
determining the collapse final states.

Using the formalism developed in
\cite{PSJ}
to study here the class of collapse models with a non-zero
tangential pressures, three main results are obtained here.
Firstly, we examine the effect of introducing small tangential
pressure perturbations, on the black hole formation process, for
a collapsing inhomogeneous dust cloud which is otherwise
pressure-free. It is seen that
a matter cloud that was collapsing to a black hole final
state could change its course, and instead result in a naked singularity
final configuration, when small tangential pressure perturbations
are allowed in an otherwise pressure-free cloud. It is seen
that the opposite scenario occurs as well.
Secondly, we generalize here earlier results on gravitational collapse with
pressure, in the presence of a non-zero cosmological constant. Finally,
from the analysis of some collapse models,
some interesting insights can be obtained on the genericity or
otherwise of naked singularity formation in gravitational collapse,
in the presence of a non-zero tangential pressure. Using the approach
developed here, it is clearly
seen how the black hole and naked singularity final states in
collapse are distinguished, depending on the initial configurations
of the matter density, the tangential stresses profiles and
the velocities of the collapsing shells. Given this initial data,
the Einstein equations then fully determine the allowed evolutions
that will lead to the formation either of a black hole or
a naked singularity.

Various subcases of this class
of collapse models have been studied in the past by different authors,
as we point out below, but our treatment here gives a unified and
complete picture on collapse final states in terms of black hole and naked
singularity formation, thus generalizing the earlier results.
Thus the role of tangential stresses towards determining the final
state of a collapsing matter cloud emerges transparently in our treatment.
This helps us to see in a collective manner the different results
in this class of collapse models, which are otherwise somewhat
scattered in a variety of approaches that do not
always make it straightforward to see the overall conclusions
in totality. Within this framework, the known results are
easily recovered, and new scenarios and insights are obtained.

In section \ref{einstein} we review as needed
the general formalism for gravitational collapse in the presence
of pressures, as restricted to the case of non-vanishing tangential
pressure and vanishing radial pressure. The set of
Einstein equations that governs the dynamical evolution
of the collapsing cloud and the procedure to solve them is outlined.
The regularity and physical reasonableness conditions that one would
like to impose upon the matter cloud, in order to avoid any unphysical
behaviour of the matter fields are described.
The initial and boundary conditions that determine the dynamical
evolution of the matter cloud are also discussed.
Section \ref{collapse} discusses the evolution of collapse,
a function related to the tangent to the outgoing null geodesics at
the singularity is evaluated, and it is the sign of this
quantity that governs the final outcome of collapse in terms
of either a black hole or naked singularity.
The occurrence of trapped surfaces during collapse is
discussed, and the visibility of the singularity is studied
through the analysis of outgoing radial null geodesics.

The formalism is then applied in section \ref{perturb}
to examine the effect of introducing small tangential pressures
on the black hole formation process, for an otherwise pressure-free
inhomogeneous dust cloud collapse.
In the section \ref{generic}, we discuss the
genericity of black holes and naked singularities obtained
as collapse endstates, for this class of collapse models.
The homogeneous and inhomogeneous dust cases are also briefly
discussed, giving a review of known results for dust, obtained as
a special case of the above treatment.
In section \ref{lambda},
we generalize some earlier results on collapse with tangential
pressure in the presence of a cosmological constant, which
is asymptotically de-Sitter or anti-de Sitter spacetime.
We study how the presence of a cosmological term in the equations
affects the dynamical behaviour of the cloud.
The study is then applied to the classes studied in
section \ref{perturb}, by adding a
cosmological term and analyzing how the dynamics is altered.
This constitutes a new class of solutions whose behaviour
was not studied earlier.
Finally, in section \ref{remarks}, we outline
the interesting features of the above treatment in
connection with previous results obtained by
other authors.

\section{Einstein Equations and Regularity conditions}\label{einstein}

The general treatment to obtain the evolution and
final fate of a spherical gravitational collapse with
non-zero pressures was developed in
\cite{PSJ},
and it constitutes the backbone upon which the present work is developed. One of the
advantages of this approach is that it shows the final stages
of collapse in a rigorous and yet
transparent fashion, where all the crucial elements that enter towards
the determination of the
final outcome are enclosed in a single straightforward equation.
It is clear that in order
for the singularity to be visible, some outgoing nonspacelike
geodesics must come
out of it to reach the faraway observers in the universe.
This visibility of outgoing geodesics
is shown to be related to the tangent to the null geodesics
near the singularity, which in turn
is shown to be dependent on the mass distribution, the pressure
and the velocity profiles for the collapsing cloud.
Once these quantities are set at the initial time, the evolution,
as governed by the
Einstein equations, determines entirely the final
outcome of collapse.

In spherical coordinates, the most general line element
for a spherically symmetric collapsing cloud of matter depends
upon three functions $\nu, \psi$ and $R$ of the comoving radial
coordinate $r$ and the comoving time $t$, and it can be
written as,
\begin{equation}\label{metric}
    ds^2=-e^{2\nu(t, r)}dt^2+e^{2\psi(t, r)}dr^2+R(t, r)^2d\Omega^2 \; .
\end{equation}
We shall assume here through out that the radial pressure vanishes.
The energy-momentum tensor, expressed in co-moving coordinates then
takes the form,
\begin{equation}
    T_t^t=-\rho; \; T_r^r=0; \; T_\theta^\theta=T_\phi^\phi=p_\theta \; .
\end{equation}
The Einstein equations then relate the three functions $\nu, \psi, R$ to the
energy density and pressures of the system, and they can then be written as,
\begin{eqnarray}\label{p}
p_r&=&-\frac{\dot{F}}{R^2\dot{R}}=0 \; ,\\ \label{rho}
\rho&=&\frac{F'}{R^2R'} \; ,\\ \label{nu}
\nu'&=&\frac{2p_\theta R'}{\rho R} \; ,\\ \label{G}
2\dot{R}'&=&R'\frac{\dot{G}}{G}+\dot{R}\frac{H'}{H} \; ,\\ \label{F}
\frac{F}{R}&=&1-G+H \; ,
\end{eqnarray}
where the function $F$ is the Misner-Sharp mass of the system and
we have introduced the functions $H(r, t)$ and $G(r, t)$, which are
defined as,
\begin{eqnarray} \label{defH}
H&=&e^{-2\nu(r, v)}\dot{R}^2 \; , \\ \label{defG}
G&=&e^{-2\psi(r, v)}R'^2 \; .
\end{eqnarray}
Since there is a scaling gauge freedom available for
the radial coordinate $r$, we can define without any loss of
generality a scale so that at the initial time we have $R(r, t_i)=r$.
To this purpose, we introduce the scaling function $v(r, t)$
such that,
$$
v(r, t_i)=1 \; .
$$
The gravitational collapse condition is given here by $\dot R < 0$.
The area function $R(r, t)$ can then be expressed as,
\begin{equation}\label{v}
    R=rv
\end{equation}
and the requirement of collapse is fulfilled once $\dot{v}<0$.

The singularity is then achieved by reaching $v=0$ (which corresponds
to $R=0$), and it is realized along the
singularity time curve $t_s(r)$. Note that as we approach the
regular center of coordinates
$r=0$, when $v\neq 0$, the energy density remains bounded, although the
physical radius $R$ goes to zero.
Therefore, this indicates that the genuine curvature
singularity, where the physical
radii of all the matter shells reach a vanishing value and the energy density
diverges, is reached only in the case
when $v=0$. This is reflected in the scaling function by the requirement
that $v(r, t_s(r))=0$ at the singularity. To avoid the
presence of shell-crossing
singularities (that can be shown to be gravitationally weak
and thus possibly removable
\cite{cross}),
we further assume that,
\begin{equation}
R'(r,t) > 0 \;.
\end{equation}
The shell focusing singularity at $v=0$ is a true curvature
singularity, and therefore
is not removable from the spacetime, and its visibility
or otherwise is the main issue
that concerns us here.

From equation \eqref{p}, it follows that the Misner-Sharp mass
$F$ in this case does not depend on time ($\dot F = 0$), and therefore it
must be a function of $r$ only. We are then left with four equations
in the six unknown $\rho(r, t),\; p_{\theta}(r, t), \; R(r,t),\;
F(r),\; G(r,t),\; H(r,t)$.

Collapse models with such a vanishing radial pressure,
but non-zero tangential stresses were studied in the mass-area
coordinates by Magli
\cite{Magli},
where the analytic structure of these solutions was derived.
The algebraic equation which governs the nature of the central singularity
was written and its dependence upon the initial configuration and the
equation of state for the matter field was analyzed. Still,
given the non-trivial
nature of the problem, the task to study the outcome of collapse in
full generality proved to be non-attainable.
Further work by Gon\c{c}alves, Jhingan and Magli
\cite{Gonc-Jhin-Mag}
studied the final stages of collapse in the limited case of
marginally bound collapse while the analysis of one of us
\cite{press}
concentrated on models with some restrictions on the
pressure profiles (namely, assuming for the metric function
$\nu=\nu(R)$). Also, the work by Harada, Nakao and Iguchi
\cite{Nakao}
analyzed the nakedness and strength of the central singularity
obtaining some general conditions which were later specialized to the
dust and `Einstein cluster' scenarios. In this work the nakedness and
curvature strength of the shell focusing singularity was studied in detail,
providing a classification of the curvature strength of the singularity
based on some conditions (namely, the strong curvature condition,
gravity dominance condition, and the limiting focusing condition).

Another interesting model with vanishing radial stresses that
has been studied is the so called `Einstein
cluster'. It is a specific subclass of models with tangential stresses
that is constituted by a collapsing cloud of counter rotating
particles. Since the tangential stress is due to the particles'
angular momentum, the `Einstein cluster' is a particularly interesting
model, because it reproduces the effects of rotation without departing
from spherical symmetry. The analysis of the singularity forming
in the `Einstein cluster'
\cite{cluster}
shows once more how departure from dust models to include pressures
does not forbid the formation of naked singularities.

With the definitions \eqref{v}, \eqref{defH},  we substitute the unknowns
$R, H$ with $v, \nu$. The definition of $F(r)$ provides one of the
two free functions
and determines the energy density via equation \eqref{rho}.
We can choose $\nu$ as the
second free function and therefore evaluate $p_\theta$
through equation \eqref{nu},
which can be written in the form,
\begin{equation}\label{ptheta}
    p_\theta=\frac{1}{2}\rho R \frac{\nu'}{R'} \;.
\end{equation}
Without any loss of generality, we can obtain $G$ from \eqref{G}
once we rewrite it as,
\begin{equation}\label{Gdot}
\dot{G}=2\nu'\frac{\dot{R}}{R'}G
\end{equation}
and we define a suitable function $A(r, v)$ from
\begin{equation}\label{At}
    A_{,v}(r, v)=\nu'\frac{r}{R'} \; .
\end{equation}
Then equation \eqref{Gdot} can be integrated to obtain,
\begin{equation}
    G(r,t)=b(r)e^{2A(r, v)} \; .
\end{equation}
The arbitrary function $b(r)$ above that results from the integration
can be interpreted following the analogy with dust LTB models.
Then $b(r)$ turns out to be
related to the velocity of the collapsing shells (more precisely
to the velocity of the infalling particles at the boundary of the cloud),
and once we write it as
\begin{equation} \label{b}
    b(r)=1+r^2b_0(r) \; ,
\end{equation}
we can see that the values of $b_0=const.$ in the dust limit correspond
to the bound ($b_0<0$), unbound ($b_0>0$) or
marginally bound ($b_0=0$) Lemaitre-Tolman-Bondi
models.

Writing equation \eqref{F} in the form of a potential,
\begin{equation}
    \dot{R}^2=e^{2\nu}\left(\frac{F}{R}+G-1\right)=U(r,t) \; ,
\end{equation}
it is possible to analyze the dynamical behaviour of different pressure
models with the usual phase space diagrams of
classical mechanics. Some examples will be discussed in section \ref{lambda}.
It turns out that depending on the velocity profile and the equation describing
the tangential stresses, collapsing scenarios and bouncing
scenarios are both allowed
\cite{Magli}.
This marks a considerable difference with the LTB dust case where
the potential $U(r,t)$ is always decreasing, thus
allowing only collapse to occur.

Finally, from equation \eqref{F} we get the equation of motion:
\begin{equation}\label{vdot}
    \dot{v}=-\frac{e^{\nu}}{r}\sqrt{\frac{F}{R}+G-1} \; ,
\end{equation}
where the negative sign has been considered in order to describe
gravitational collapse. In some cases equation \eqref{vdot} can be integrated
to obtain the function $v(r, t)$, thus solving completely the system of
Einstein equations. In other situations, considerations on
the behaviour of $v(r,t)$
near the center of the collapsing cloud suffice to provide
the analysis of the dynamical
behaviour of the system approaching the singularity.


A viable physical description of the gravitational collapse
of a massive body must be subdued to certain
constraints and regularity conditions in order to ensure that the
physical quantities that
appear in the model are well defined and regular everywhere.

Requiring regularity of the energy density at the initial surface from which
the collapse commences, and at the center of the cloud throughout
collapse, imposes some constraints on the classes of mass profiles that
can be considered. Specifically, the mass profile $F(r)$ must have
the form,
\begin{equation}\label{mass}
    F(r)=r^3M(r) \; ,
\end{equation}
which immediately implies the regularity of the density function
at the center of the cloud at all regular epochs. By this prescription,
we ensure that the Misner-Sharp
mass $F(r)$ vanishes at the center of the cloud and that the energy density
$\rho$ away from the singularity
has a finite value at the center. A further condition that
may be imposed sometimes is that the energy density does not present a cusp
at $r=0$, and that it is a smooth function of the radial coordinate $r$.
This can be achieved by requiring that the linear term of the energy density
expansion near the center vanishes, which is reflected in the
requirement that $M'(r)=0$ at $r=0$. Of course, this is an additional
requirement and in a general physically realistic scenario, this may or may
not be realized. In any case, requiring smoothness brings in some
additional mathematical simplification.

The requirement that there is no pressure gradient at $r=0$
implies that the tangential pressure must become equal to the radial pressure
(i.e. zero in our case) at the center.
This imposes the condition that $p_\theta$ goes
as $r$ at least, near the center, and implies regularity
for the metric function $\nu$
via equation \eqref{nu}. This condition then implies that we
have in general,
\begin{equation}\label{nu-def}
    \nu(r,v)=r^2g(r, v) \; ,
\end{equation}
with $g$ being a regular function and $\nu(r, t)=\nu(r, v(r, t))$.
Regularity of $\nu$ implies regularity of the function
$A$ via equation \eqref{At},
and we have,
\begin{equation}\label{A-def}
    A(r, v)=r^2a(r, v) \; ,
\end{equation}
with $a$ being again a regular function.
The relation between $a$ and $g$ is given by the equation \eqref{At},
which can be written as
\begin{equation}\label{a}
  a_{,v} =\frac{2g+rg'}{R'} \; .
\end{equation}

As we have seen before, a specific behavior of $R$ is required in order to
avoid shell crossing singularities. The choice of $R$ satisfying $R'>0$ implies
further regularity for the energy density.
This in turn gives some constraints on the non-vanishing terms in the expansion
of the energy density at a constant time surface. Namely, we find that in order
to avoid the shell crossing singularities, we must require $M'\leq 0$, which,
from a physical point of view, is a reasonable condition since we expect
the density to be higher at the center of the cloud, and decreasing
as we move away from the center.
In the case when $M'(0)=0$, $\rho$ turns out to be a
decreasing function from the center outwards if $M''(0)<0$.

As noted, from the regularity of $\dot{v}$ at the center of the
cloud, we get the behaviour of $b$ as in equation \eqref{b},
and this turns out to
be related to the velocity profile, and thus the kinetic energy,
of the infalling particles.
Positivity of $G$ is thus reflected in the condition that
\begin{equation}\label{b0}
    b_0\geq -\frac{1}{r^2} \; .
\end{equation}
Furthermore, the requirement that $R'>0$ implies that moving away
from the center, the contribution of the kinetic energy
to the total energy of the collapsing cloud decreases
\cite{cross}.
Therefore, the requirement of avoidance of shell crossing implies that
some further condition must be imposed upon $b_0(r)$.
In general, from the Misner-Sharp mass we get
\begin{equation}
    F'(1-G+H)>F(H'-G') \; ,
\end{equation}
which, in the limit of
the dust case, becomes,
\begin{equation}
    b_0M'> Mb_0' \; .
\end{equation}

The marginally bound case ($b_0=0$) is obviously allowed,
and we see that considering only the constant values for the function $b_0(r)$
(which correspond in the dust case to the bound and unbound LTB models),
the bound case, i.e. $b_0<0$ might imply the occurrence of
shell crossing if $M'>0$.
The bound case is therefore allowed under the restrictions imposed as
above. We note from equation \eqref{b0}
that the permitted values for $b_0$ have a limited range,
namely $b_0\in (-\frac{1}{r_b^2}, 0)$,
where $r_b$ is the boundary of the cloud (an infinite cloud cannot
therefore have a bound velocity profile).
On the other hand, the unbound models, which are also allowed,
describe a collapse scenario
that is not uniquely gravitational. Some additional
energy is impressed to the
infalling particles from outside, causing the particles
to have a non zero velocity
at all times. We note that in the unbound case, solutions with $\dot{R}=0$
at the initial time are not possible.

In the more general situation where $b_0=b_0(r)$, the fact that certain
velocity profiles are allowed or not might depend on the boundary conditions
imposed on the matter cloud. Therefore, for some specific cases,
both positive and
negative functions $b_0(r)$ can be considered, without
causing the presence of shell crossings.

At this point, we note that the requirement
that the regularity conditions must be satisfied
constrains the class of allowed models and the
allowed equation of state. Regularity would be reasonable
to demand from the point of view of the physical validity,
since it guarantees that the metric functions and physical
quantities are well behaved at any regular epochs of collapse,
and at the center. We see that this imposes certain
non-trivial restrictions on the possible choice of the
matter models. In fact, we notice how regularity and the
choice of a linear equation of state of the type $p_\theta=k\rho$,
as was assumed in some previous works (see for example,
\cite{Singh}),
are mutually exclusive.
This is immediately seen from equation \eqref{nu}
which now becomes,
\begin{equation}
    \nu'=2k\frac{R'}{R},
\end{equation}
and once integrated and evaluated at the
initial time $t=t_i$ it gives,
\begin{equation}
    \nu_i(r)=2k\ln r+const.
\end{equation}
Thus the metric function diverges as $r$ tends to zero,
while on the other hand regularity demands that $\nu\sim r^2$
at all times, near the center.
We thus see that in comoving coordinates the
assumption of a linear equation of state leads to
a breakdown of the coordinate system at the center,
or the metric functions not being regular.
This would imply that the initial data set is not
regular, or that a more suitable coordinate system is
required. The inspection of the Kretschmann scalar
$K$ for this model shows that for any $k\neq 0$,
the central shell presents a curvature singularity
(where $K$ diverges) at all times, and thus the
initial data surface is not regular. This is confirmed
also by the divergence at $r=0$, of the energy density
$\rho$, as evaluated from equation \eqref{rho}.
It is seen that at the initial epoch also, the
energy density diverges necessarily at the center.
For the tangential pressure models, the choice
$p_{\theta}=k \rho$ turns
out to be consistent with the regularity
condition $F(r)=r^3M(r)$ only in the dust case,
where $k=0$.

In order for the matter model to be physically realistic, suitable
energy conditions, in particular the weak energy conditions, must be satisfied.
That is, the energy density as measured by local observers must be non-negative,
and for any timelike vector $V^a$ we require,
\begin{equation}
    T_{ab}V^aV^b \geq 0 \; ,
\end{equation}
which reduces to
\begin{equation}
    \rho \geq 0, \; \rho+p_\theta \geq 0 \; .
\end{equation}

Since the physical radius $R$ is always positive,
to ensure the positivity of energy density we must require $F'$ and $R'$
or their ratio to be positive. As we have seen, positivity
of $R'$ ensures that there
are no shell crossing singularities occurring,
while the fact that $F'$ is positive
is granted by the choice of positive $M(r)$ with a suitable behaviour.
From equation \eqref{ptheta} we see that positivity of $\rho+p_\theta$ is
then given by
\begin{equation}\label{energycon}
    \frac{1}{2}\frac{R}{R'}\nu'\geq-1
\end{equation}
and is satisfied automatically (and not only) by any increasing function $\nu$.
The sign of $\nu'$ is the only factor deciding positivity
or negativity of the tangential
pressure and it is worth noting that negative tangential
stress profiles are also allowed and
they satisfy energy conditions provided that the relation expressed in
equation \eqref{energycon} is satisfied. Furthermore, for small values of $r$,
from equation \eqref{energycon} we see that regardless of the values taken by
$\nu$, there will always be a neighborhood of $r=0$ for which $|p_\theta|<\rho$
and therefore $\rho+p_\theta\geq0$ is automatically satisfied in a close
neighborhood of the center.


Finally, specifying the initial conditions for the collapsing matter cloud
consists in prescribing the values of the three metric functions and that of the
density and stress profiles at the initial time $t_i$, given as functions of $r$
\cite{Initial}.
These can be written as,
\begin{eqnarray}\nonumber
    \rho(r, t_i)&=&\rho_i(r), \; p_\theta(r, t_i)=p_{\theta_i}(r), \\ \nonumber
    R(r, t_i)&=&R_i(r), \; \nu(r, v(r, t_i))=\nu_i(r), \; \psi(r, t_i)=\psi_i(r).
\end{eqnarray}
As noted earlier, at the initial time we have taken the scale
such that $R_i=r$, which implies that the scaling function $v=1$
at the initial epoch.
Furthermore, from $R'_i=1$ we get $v'(r, t_i)=0$.

Not all of these initial value functions can be chosen arbitrarily.  In fact,
the scaling function, the choice of the mass profile, and the Einstein equations
impose some constraints. From equations \eqref{rho} and \eqref{mass},
we see that
\begin{equation}
    \rho_i=\frac{F'(r)}{r^2}=3M(r)+rM'(r) \; ,
\end{equation}
while from equation \eqref{ptheta} we get,
\begin{equation}
    p_{\theta_i}=\frac{1}{2}\nu'_i(r)\rho_i(r)r \; .
\end{equation}
Also, from equation \eqref{nu-def} we can write,
\begin{equation}
    \nu_i(r)=r^2g(r, v(r, t_i))=r^2g_i(r) \; .
\end{equation}
In turn, $\nu_i$ can be related to the function $a$, defined by equation
\eqref{A-def}, at the initial time via equation \eqref{At},
\begin{equation}
    a_{,v}(r, v(r, t_i))=(a_{,v})_i=2g_i+rg_i' \; .
\end{equation}
Finally, we can relate the initial condition for $\psi$ to the initial value
of the function $A(r, t)$ through equation \eqref{defG} and equation \eqref{At},
\begin{equation}
    A(r, v(r, t_i))=A_i(r)=-\psi_i-\frac{1}{2}\ln b \; .
\end{equation}

If we want the initial collapse configuration to be not trapped,
in order to see how different initial matter configurations evolve forming
trapped surfaces, we must further require,
\begin{equation}
    \frac{F(r_b)}{R_i}=r_b^2M(r_b)< 1 \; ,
\end{equation}
which will impose some restrictions on the possible
choices of the radial boundary in accordance with
$M(r)$. This condition reflects on the initial configuration for $G$ and $H$
since $\frac{F}{R}<1$ implies $H<G$. Therefore to avoid trapped surfaces at the
initial time the velocity of the infalling shells must satisfy
\begin{equation}
    -\dot{R}>\sqrt{b}e^{A+\nu} \; .
\end{equation}
This means that the scenario where collapse commences from an
equilibrium configuration where $\dot{R}=0$,
can be taken only as a limiting case.

As we noted above, considering only the presence of the tangential
pressures implies conservation of the Misner-Sharp mass during the collapse.
Therefore,
matching to an exterior spherically symmetric vacuum solution leads inevitably to
consider the Schwarzschild metric
\cite{matching}.
The collapsing cloud has a compact support within
the boundary taken at $r=r_b$, and the pressure of the matter
composing the collapsing star
is assumed to vanish at the boundary. Matching conditions imply continuity of the
metric and its first derivatives across this surface, although it should be noted
that eventual discontinuities in the first
derivatives of the metric across the boundary
may be interpreted as an additional distribution of matter acting as a layer
separating the interior from the exterior. In this case,
the jump of the second fundamental
form across the boundary is related to an additional stress-energy tensor layer
confined to the boundary. Such a matching is in principle always possible, and
together with the regularity conditions and the energy conditions, might impose some
further restrictions on the initial configurations allowed.

\section{Collapse evolution and final state}\label{collapse}

As we show below, the possible future evolution of the collapsing matter
cloud depends upon the choice made for the matter content, namely the two free
functions that determine the mass profile and the tangential pressure profile,
and the initial conditions.

In order to investigate the final outcome of collapse, we will study the
time curves that lead the shell of matter labeled with a specific coordinate value
$r$ to fall into the singularity. As already mentioned, the spacetime singularity is
achieved for $v=0$, and the divergence of the energy density (and eventually of the
tangential pressure) indicates that it must be a true curvature singularity.

Given the fact that the function $v(r, t)$ is monotonically decreasing in time,
we can invert it and consider the time $t$ as a
function of the variables $r$ and $v$.
This is equivalent to considering the area-radius
coordinates $(r,R)$ as done in some earlier works
(see e.g. \cite{Magli, Nakao})
with the relation between $R$ and $v$ given by equation \eqref{v}.
The function $t(r,v)$ will identify the time at which
the shell labeled by a specific value $r$ reaches
the event $v$. In this manner, the occurrence of the singularity is described by
the time curve $t_s(r)=t(r, 0)$. The tangent of the
singularity curve $t_s(r)$ at $r=0$,
{\it i.e.} at the central singularity, is then seen to be related to the eventual
existence of outgoing radial null geodesics escaping away
from the singularity, and therefore
to the possible visibility or otherwise of the
singularity itself to faraway observers.

From equation \eqref{vdot} we can write,
\begin{equation}
dt=-\frac{e^{-\nu(r, v)}}{\sqrt{\frac{M(r)}{v}+
\frac{b(r)e^{2A(r, v)}-1}{r^2}}}dv \; ,
\end{equation}
with the minus sign chosen in accordance with $\dot{v}<0$, in order to describe
the collapse. Integrating the above equation,
while treating the radial coordinate $r$
as a constant, we get
\begin{equation}
    t(r, \bar{v})=t_i+\int^1_{\bar{v}}\frac{e^{-\nu}}
{\sqrt{\frac{M}{v}+\frac{be^{2A}-1}{r^2}}}dv \; .
\end{equation}
It follows therefore that, the time taken by the shell at any given value of $r$
to reach the singularity, located at $v=0$, is given by,
\begin{equation}
t_s(r)=t_i+\int^1_0\frac{e^{-\nu}}{\sqrt{\frac{M}{v}+\frac{be^{2A}-1}{r^2}}}dv \; .
\end{equation}

We note that $t(r, v)$ is in general a differentiable function at least
outside the singularity, which would be the case because of the regularity of all
the functions involved. That is the case in many of the known collapse
models. In any case, whenever it is at least a $C^2$
function, we can always write it as an expansion near the central shell $r=0$
in the form,
\begin{equation}\label{t}
t(r, v)= t(0, v)+r\chi_1(v)+o(r^2) \; ,
\end{equation}
with
\begin{equation}
    t(0,\bar{v})=t_i+\int^1_{\bar{v}} \frac{1}{\sqrt{\left
(\frac{M(0)}{v}+b_0(0)+2a(0,v)\right)}}dv
\end{equation}
and $\chi_1(v)=\left.\frac{dt}{dr}\right|_{r=0}$, given by,
\begin{equation}
    \chi_1(\bar{v})=-\frac{1}{2}\int^1_{\bar{v}}\left.\frac{\frac{M'}{v}+
    \frac{rb'e^A+rbA'e^A-2(be^A-1)}{r^3}}{\left(\frac{M}{v}+
\frac{be^{2A}-1}{r^2}\right)^{\frac{3}{2}}}\right|_{r=0}dv \; .
\end{equation}
Note that the above quantities are defined
only if $\frac{M(0)}{v}+b_0(0)+2a(0,v)>0$.
This may be called the reality condition for the gravitational collapse to
take place.
The choices of $b(r)$ and $A(r, v)$, made from \eqref{b} and \eqref{A-def}
imply that as $r\rightarrow 0$, we must have
$\frac{be^{A}-1}{r^2}\rightarrow b_0+2a(0,v)$,
and therefore we see how $\nu$ cannot have constant terms or terms linear
in $r$ in a close neighborhood of $r=0$. This justifies the regularity conditions
for $b(r)$ and $\nu(r)$ that were deducted earlier.
Furthermore, we can see how the regularity condition for $A$, that comes
directly from that for $\nu$, ensures that $t(0,v)$ and $\chi_1$ do not diverge.
Evaluating $\chi_1$ explicitly we get
\begin{equation}
    \chi_1(\bar{v})=-\frac{1}{2}\int^1_{\bar{v}}\frac{\frac{M'(0)}{v}+
  b_0'(0)+2a'(0,v)}{\left(\frac{M(0)}{v}+b_0(0)+2a(0,v)\right)^{\frac{3}{2}}}dv \; .
\end{equation}

By continuity, we can therefore take that the singularity curve
$t_s(r)$, which corresponds to $v=0$, is differentiable, or in any case even
if it is only a $C^2$ function, we can take it that
it has at least a well-defined tangent at the singularity. This is the case
in important known collapse models such as the Tolman-Bondi-Lemaitre
dust collapse, or the `Einstein cluster' models, and also for the
Oppenheimer-Snyder-Datt homogeneous dust collapse.
The singularity curve is computed
as the time it takes for the shell located at $r$ to reach the singularity
and can then be written as,
\begin{equation}\label{ts}
    t_s(r)= t_0+r\chi_1(0)+o(r^2) \; ,
\end{equation}
where $t_0=t_s(0)$ is the time at which the central shell becomes singular.
It can be shown that for any generic initial configuration there exist
classes of dynamical collapse evolutions that have a well defined
singularity curve.


While the collapse evolves, the trapped surfaces may form in the
collapsing cloud. If the trapped surfaces form at a time anteceding the formation of
singularity then the latter will be covered, thus giving rise to a black hole at
the end of the collapse. If, on the other hand, the trapped
surfaces do not form, or form
at a later time after the singularity, then the singularity might be visible to
faraway observer
\cite{Global aspects}.

The boundary of the trapped region is delimited by the apparent horizon.
The equation for the apparent horizon in a spherically symmetric spacetime is given
by $g^{ij}R_{,i}R_{,j}=0$, which for the metric \eqref{metric} becomes $G-H=0$. From
equation \eqref{F} we can write it as,
\begin{equation}\label{horizon}
    \frac{F}{R}=1 \; .
\end{equation}
The behavior of the apparent horizon for the collapsing cloud with
vanishing radial pressure is analogous to that of the dust case, since
$\frac{F}{R}=\frac{r^2M(r)}{v}$ in both these cases. If we note
that as $v\rightarrow 0$
we must have $r\rightarrow 0$ on the apparent horizon, it is immediately clear
that the only singularity that can eventually be visible in the present case
is that at the center of
the collapsing cloud. In fact, since it is not possible to satisfy $1-\frac{F}{R}>0$
near the singularity but away from the center,
the region surrounding the singularity cannot be timelike,
and therefore any singularity that might eventually form near the
center with $r>0$ must not be visible.

The difference with the dust case is then given by the different
possible behaviours of the function $v(r,t)$. It is this function, as given from
equation \eqref{vdot}, that determines the apparent horizon curve in dependence
of the mass, velocity and the tangential stress profiles. We can
see the apparent horizon
as a curve $r_{ah}(t)$ given by
\begin{equation}\label{ah}
    r_{ah}^2=\frac{v_{ah}}{M(r_{ah})} \; ,
\end{equation}
with $v_{ah}=v(r_{ah}(t), t)$.  Inversely, we can also write it as a curve
$t_{ah}(r)$, which represents the time at which the shell labeled as $r$ becomes
trapped.

In order to understand when the central singularity
may be visible to faraway observers in the spacetime, we shall analyze the time
curve of the apparent horizon which is given by,
\begin{equation}\label{t-ah}
    t_{ah}(r)=t_s(r)-\int_0^{v_{ah}(r)}\frac{e^{-\nu}}
{\sqrt{\frac{M}{v}+\frac{be^{2A}-1}{r^2}}}dv \; ,
\end{equation}
where $t_s(r)$ is the singularity time curve,
whose initial point is $t_0=t_s(0)$.
Near $r=0$, equation \eqref{t-ah} can be written in the form,
\begin{eqnarray}\nonumber
    t_{ah}(r)&=&t_{ah}(0)+\left.\frac{dt_{ah}}{dr}\right|_{r=0}r+o(r^2)=\\
    &=&t_0+\chi_1(0)r+o(r^2) \; .
\end{eqnarray}
From this, it is easy to see how the presence of stresses affects
the time of the formation of the apparent horizon.
In fact, all the initial configurations that cause $\chi_1(0)$ to be
positive force the boundary of the trapped surfaces to form
after the occurrence of the singularity, thus leaving the `door open' for
the null geodesics to possibly come out, away from the singularity.

We can verify the visibility of the central singularity through the
behaviour of radial null geodesics. This means that
in order to verify the visibility
of the singularity we must check if there are any
future directed null or non-spacelike
geodesics that terminate in the past at the singularity (for a detailed analysis of
outgoing geodesics in singular spacetimes originating from collapse see
\cite{Giambo}).
It can be shown that if the singularity is radially censored
then it must be completely censored.
On the other hand, if there are any future directed non-spacelike curves coming out
from the singularity and reaching external observers then the singularity is naked.
Further, in
\cite{PSJ}
it was proven that $\chi_1(0)>0$ is a necessary and sufficient condition for outgoing
radial null geodesics to escape the singularity, at least locally.
Outgoing radial null geodesics are given by the equation,
\begin{equation}\label{geodes}
    \frac{dt}{dr}=e^{\psi-\nu}=\frac{R'}{\sqrt{b}}e^{-\nu-A} \; ,
\end{equation}
with the positive sign considered in order for $t$
to be increasing moving away from the singularity.

Now, introducing the variable $u=r^\alpha$,
we can write equation \eqref{geodes} as
\begin{eqnarray}\nonumber
    \frac{dR}{du}&=&\frac{1}{\alpha r^{\alpha-1}}\left(R'+\dot{R}e^{\psi-\nu}\right)= \\ \label{geodesic}
    &=&\frac{R'}{\alpha r^{\alpha-1}}\left(1-\frac{e^{-A}}{\sqrt{b}}
\sqrt{\frac{F}{R}+G-1}\right) \; .
\end{eqnarray}
We thus look for a radial null geodesic terminating in the past
at the singularity, whose equation near $r=0$ takes the form
\begin{equation}
    R=x_0u,
\end{equation}
with $x_0$ finite and positive. Such null geodesics terminate
at the singularity with a definite tangent and correspond to the curve
\begin{equation}
    t(r)=t_0+x_0r^\alpha \; .
\end{equation}
For a specific choice of $\alpha=\frac{5}{3}$, we can
solve equation \eqref{geodesic} and evaluate the value of the tangent
to the null geodesics at the central singularity which is given by $r=0$, $t=t_0$.
Thus obtaining,
\begin{equation}\label{x0}
    x_0=\lim_{r\rightarrow 0}\lim_{t\rightarrow t_0}
\frac{R}{u}=\frac{dR}{du}\left|\frac{}{}_{(0,t_0)}\right.=
\left(\frac{3}{2}\sqrt{M_0}\chi_1(0)\right)^{\frac{2}{3}} \; ,
\end{equation}
which is positive and finite at the central singularity if $\chi_1(0)>0$.
It is now clear how $\chi_1$ is directly related to the tangent to the null geodesics
at the singularity. In this case we conclude that the future directed
radial null geodesics do come out of the singularity to reach far away observers.


Let us stress that it is the sign of the quantity $\chi_1(0)$
that uniquely determines
the visibility of the singularity, as in fact $\chi_1(0)>0$
implies that the time curve
of the apparent horizon is future directed increasing, and future directed null
geodesics in that case do come out. Therefore the central singularity
occurring at $t_0$ is not covered.

All those initial configurations for which $M'(0)$, $b_0(0)$ and
$a_1(v)$ cause $\chi_1(0)$ to be positive will force the apparent horizon
to appear after the formation of the
singularity, and will allow the null geodesics to come out of the singularity
that forms at the center of the collapsing cloud.
We conclude that the condition $\chi_1(0)>0$
not only implies that the trapped surfaces form at a later stage during
collapse than the singularity, but it is also the sufficient condition
for null geodesics from the singularity to be visible
(at least locally) to external observers.

\section{Small pressure perturbations in dust collapse}\label{perturb}

We would now like to examine here, using the formalism
above, how the final fate of the gravitational collapse is
affected when pressure perturbations are introduced in a
pressure-free inhomogeneous dust cloud, which was otherwise
going to terminate in a black hole final state. This provides
a useful insight into the role of pressure in gravitational
collapse dynamics. Certain classes of collapse models
with a non-zero tangential pressure are analyzed, their properties
are discussed, and the collapse end-states are examined.

For the sake of definiteness, we consider the case
when the Misner-Sharp mass is given by an expression near
the center as,
\begin{equation}
    M(r)=M_0+M_1r+M_2r^2+o(r^2).
\end{equation}

Further expressing the function $g(r,v)$
(and therefore the tangential pressure)
in the form,
\begin{equation}
    g(r,v)=g_0(v)+g_1(v)r+g_2(v)r^2+ o(r^2) \; ,
\end{equation}
we have,
\begin{equation}\label{pt}
p_\theta=\frac{r^2}{vR'^2}\left(3M_0g_0+4M_1g_0r+\frac{9}{2}M_0g_1r+...\right) \; .
\end{equation}
With these expressions, we are then able to write the quantity $\chi_1(0)$
in dependence of the physically relevant profiles
$M(r)$, $b(r)$, $a(r, v)$ as,
\begin{equation}\label{chi}
    \chi_1(0)=-\frac{1}{2}\int^1_{0}\frac{\frac{M_1}{v}+
    b_0'(0)+2a_1(v)}{\left(\frac{M_0}{v}+b_0(0)+2a_0(v)\right)^{\frac{3}{2}}}dv \; ,
\end{equation}
where the function $a(r,v)$ (which is related to $\nu$ via equation \eqref{At})
has been written as,
\begin{equation}
    a(r,v)=a_0(v)+a_1(v)r+a_2(v)r^2+o(r^2) \; .
\end{equation}

We see now immediately that the behaviour of the tangential
pressure (be it positive, negative, increasing or decreasing) is here
reflected in the terms $a(0,v)$ and $a'(0,v)$ in $\chi_1(0)$, and can influence
the final outcome of the gravitational collapse.
In fact, it is the choice of the coefficients
of the initial density, velocity, and stress profiles that determines the quantities
$M$, $b$ and $a$, which are responsible for the behaviour of $\chi_1(0)$, which
in turn determines the tangent to the singularity curve at the origin.
As we noted earlier, the sign of $\chi_1(0)$ uniquely determines
the final fate of collapse in
terms of either a black hole or a naked singularity. As it was shown in
\cite{PSJ},
positivity of $\chi_1(0)$
is a sufficient condition for outgoing null geodesics to come out from
the singularity, thus making it visible to external observers.
In some cases of physical interest it turns out that $\chi_1(0)=0$.
This is likely to happen when marginally
bound collapse with only quadratic terms in the expansion of
energy density and pressure is considered, but it is not the only possibility.
In general, whenever $\chi_1(0)=0$, the analysis of higher order terms must
be carried out. In this case, we can write $t_s(r)$ as
\begin{equation}
    t_s(r)=t_0+r^2\chi_2(0)+o(r^3)
\end{equation}
and evaluate $\chi_2(0)=\left.\frac{d^2t}{dr^2}\right|_{r=0}$ as
\begin{eqnarray}\nonumber
    \chi_2(0)&=&\int^1_0\left[\frac{3}{8}\frac{(2a_1+b_0'(0))^2}
{\left(\frac{M_0}{v}+b_0(0)+2a_0\right)^\frac{5}{2}}+\right.\\ \nonumber
    &&-\frac{1}{2}\frac{\frac{M_2}{v}+2a_2+2a_0^2+2b_0(0)a_0+b''_0(0)}
    {\left(\frac{M_0}{v}+b_0(0)+2a_0\right)^{\frac{3}{2}}}+ \\ \label{chi2}
    &&-\left.\frac{g_0(v)}{\sqrt{\frac{M_0}{v}+b_0(0)+2a_0}}\right]dv \; .
\end{eqnarray}

We also note here that
the negative tangential pressures are not forbidden by the energy conditions.
Here we observe, how the presence of negative pressures,
that may be due to the presence of exotic matter in the form of
dark matter or some other repulsive effects, may influence the final
stages of collapse,
since it will affect the value of $\chi_1$ through the function $a_1(v)$ given
in \eqref{a}.

The second order term in the equation for the singularity curve
is given by equation \eqref{chi2}.
It turns out that this term becomes relevant in many physically realistic
scenarios which consider the mass profile and the density
profile to be written with only quadratic terms in $r$.
From the regularity conditions analyzed
earlier, we can further see that $M'(0)=0$ (that is reflected
in the requirement that $M_1=0$) is a reasonable assumption.
This is consistent with the analysis of energy density profiles, that, for physically
realistic scenarios, have only quadratic terms in the expansion.
The energy density $\rho$, can thus be written
as $\rho(r,t)=\rho_0(t)+\rho_2(t)r^2+\rho_4(t)r^4+...$.

In the situation where mass and
density profiles are given only in quadratic terms, we then have,
\begin{equation}
M_{2n+1}=a_{2n+1}=g_{2n+1}=0 \; .
\end{equation}

We can now examine a few physically
relevant classes of collapse models with such a property,
together with the prescription for marginally bound collapse ($b_0=0$).
This choice of the mass, pressure profile and velocity profile
causes $\chi_1(0)$ to vanish and the final outcome of collapse is decided
by the $\chi_2(0)$ term, which in this case reads,
\begin{equation}\label{chi2quad}
\chi_2(0)=-\frac{1}{2}\int_0^1 \frac{\frac{M_2}{v}+2a_2+2a_0^2+2g_0
\left(\frac{M_0}{v}+2a_0\right)}{\left(\frac{M_0}{v}+2a_0\right)^{\frac{3}{2}}}dv.
\end{equation}
The analysis is then carried out along the same lines described above.

Firstly, an important class of collapse models,
also to be noted as the reference frame, is that of an
inhomogeneous dust cloud. Inhomogeneous
dust has been widely studied and the occurrence of naked singularities
in such cases is well-known. From the framework given above, we recover the
inhomogeneous dust when $a(r,v)=g(r,v)=0$, and that together with
the requirement that only quadratic terms appear in the density, yields
\begin{equation}
\chi_2^{dust}(0)=-\frac{1}{2}\int_0^1 \frac{M_2\sqrt{v}}{M_0^{\frac{3}{2}}}dv
=-\frac{1}{3}\frac{M_2}{M_0^{\frac{3}{2}}} \; .
\end{equation}
The collapse will end in a naked singularity for all negative
values of $M_2$. Further, in this case, the structure of the apparent horizon
can be evaluated explicitly. Note that for $M(r)=M_0$ the above case
reduces to the Oppenheimer-Snyder-Datt collapse scenario,
describing homogeneous dust, and all the terms $\chi_i$s vanish
as the singularity is simultaneous.

Let us now consider the effect of introducing small
tangential pressure perturbations in an otherwise pressure-free
inhomogeneous dust cloud collapse, that was going to a black hole
final state. To this aim, we choose a class of collapse models
restricted by the assumption that the second order term in the
mass profile vanishes. We see that if $M_2=0$ then
$\chi_2^{dust}(0)=0$, and the final outcome of collapse for
the inhomogeneous dust cloud is then decided by the next order term,
namely $\chi_4^{dust}(0)=-\frac{1}{3}\frac{M_4}{M_0^{\frac{3}{2}}}$.
Exactly in the same manner as described above, all positive
values of $M_4$ will cause the collapse to end in a black hole.
If now we introduce an arbitrarily small tangential pressure
of the form \eqref{pt}, we see that near $r=0$ this can cause the
collapse to end in a naked singularity, whenever the function
$g_0(v)$ is chosen in such a way that $\chi_2(0)$, as given by
equation \eqref{chi2quad} with $M_2=0$, is positive. We
therefore have constructed a class of models of small tangential
pressure perturbations of the LTB collapse scenario that
drastically changes the final outcome of collapse on
introduction of small pressures.

A second illustrative class of models is
that of a `quasi-Hookean' pressures, which is given by a specific
choice of the tangential stresses corresponding to
\begin{equation}
    a_{2n-2}=(-1)^{n+1}\mu_0\left(1-\frac{1}{v^2}\right) \; ,
\end{equation}
for $n=1, 2, 3...$.
This choice leads to an energy function $G(r,v)$ of the form
\begin{equation}
    G(r,v)=\frac{1+b_0r^2}{\left[1+\mu_0\left(1-\frac{1}{1+r^2}\right)
\left(1-\frac{1}{v^2}\right)\right]^2} \; .
\end{equation}
In this case, putting again $b_0=0$, we get
\begin{eqnarray}
\chi_2(0)&=&-\frac{1}{2}\int^1_0 \frac{M_2v^2+2M_0\mu_0}
{\left(M_0v+2\mu_0v^2-2\mu_0\right)^{\frac{3}{2}}}dv+\\ \nonumber
&+&\int^1_0\frac{\mu_0v\left(v^2-1\right)\left(1-\mu_0\left(1+
\frac{1}{v^2}\right)\right)}
{\left(M_0v+2\mu_0v^2-2\mu_0\right)^{\frac{3}{2}}}dv \; .
\end{eqnarray}
Some more considerations on the dynamical behaviour of
this particular case will be presented in section \ref{lambda} also.
What we can see from the above is that the values of the
constants $M_2$ and $\mu_0$ are responsible for the final outcome
of collapse in terms of black hole or naked singularity, and
that either of the outcomes are possible depending on
the choices of these parameters.

Finally, it is interesting to study the behaviour of
a subclass of the previous model, choosing the class of small tangential
stresses in the form,
\begin{equation}
    p_\theta=6\frac{M_0g_2}{v^3}r^4+o(r^6) \; .
\end{equation}
In this case we see that $a_0=a_1=0$, and near $r=0$ we have
$A(r,v)=a_2(v)r^4+o(r^6)$ and $g_2=\frac{v}{4}a_{2,v}$. Then the final
outcome of collapse is decided by the sign of
\begin{eqnarray}\nonumber
\chi_2(0)&=&-\frac{1}{2}\int^1_0 \frac{\sqrt{v}(M_2+2a_2v)}
{M_0^{\frac{3}{2}}}dv=\\ \label{example}
&=&\chi_2^{dust}(0)-\frac{1}{M_0^{\frac{3}{2}}}\int^1_0a_2v^{\frac{3}{2}}dv \; .
\end{eqnarray}
We see that in this case it is possible that the introduction
of a suitable tangential stress in the inhomogeneous dust cloud
will cause it to end in a naked singularity, while the same model
with a vanishing pressure resulted in a black hole, and viceversa.
The condition to be satisfied by the tangential stress
is $3\int_0^1a_2v^{\frac{3}{2}}dv<-M_2$.

To see this more explicitly,
in analogy with the previous example, consider for instance
\begin{equation}
a_2=-\mu_0\left(1-\frac{1}{v^2}\right) \; ,
\end{equation}
then the tangential stress near the center of the cloud will be
given by,
\begin{equation}
p_\theta=3\frac{M_0\mu_0}{v^5}r^4+o(r^6).
\end{equation}
We note that the tangential pressure as well as the density
diverge at the singularity. The collapse here ends in a naked singularity
whenever
\begin{equation}\label{exampleM}
M_2<-\frac{24}{5}\mu_0 \; ,
\end{equation}
as can easily be calculated from equation \eqref{example}.

\begin{figure}[hh]
\centering
\includegraphics[scale=0.4]{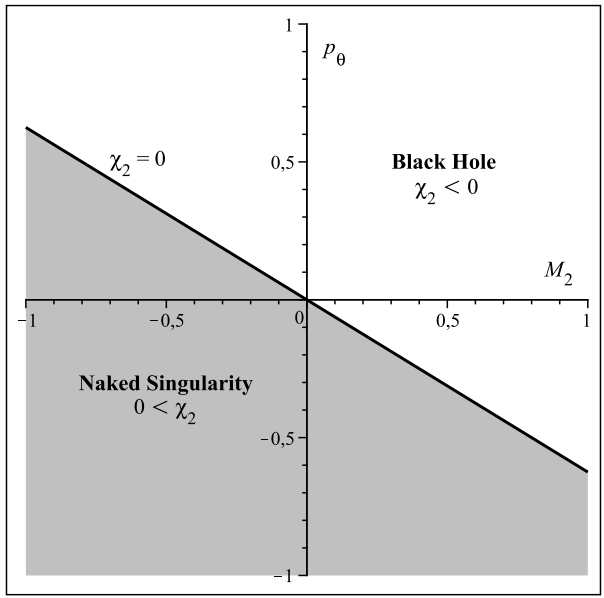}
\caption{The Naked singularity region and the black hole region in the parameter space $\{M_2,p_\theta\}$ for fixed values of $M_0$, $r$ and $v$.}
\label{Fig}
\end{figure}

In Fig. \ref{Fig}, the occurrence of black hole
and naked singularity phases in the latest example are shown,
depending on the possible values chosen in the parameters space of
$p_\theta$ and $M_2$, for fixed values of $M_0$, $r$ and $v$
(taken at the initial time). The line $\chi_2=0$ corresponds to
$p_\theta=-\frac{5}{8}\frac{M_0M_2}{v^5}r^4$, and acts as a critical
surface separating the two outcomes, in which case the evaluation
of higher order $\chi_i$ is necessary in order to determine
the final fate of collapse. The axis $p_\theta=0$ corresponds to
the LTB inhomogeneous dust case, with the origin being the
Oppenheimer-Snyder-Datt model.

We see clearly that to have a naked singularity in the dust case,
we must require $M_2<0$. This is not the case anymore in the more general
situation with non-vanishing tangential stresses.
In fact, such scenarios allow for
positive values of $M_2$ to lead to a naked singularity, provided that the
tangential stress (and hence $\mu_0$ in this example) is negative,
as well as negative values of $M_2$ leading to black holes whenever
a suitable positive tangential pressure is introduced. The interesting
feature here is that the introduction of tangential pressures
does not rule out the possibility of formation of
naked singularity.

This shows that the conjecture suggested in
\cite{Magli}
does not hold (even in the more restrictive case
of quadratic density and pressure profiles), and that naked singularities are
a general feature of gravitational collapse with pressures.
The conjecture stated that given a known Lemaitre-Tolman-Bondi
collapse model ending in a black hole, all quadratic pressure
perturbations to the model will necessarily originate a black hole.
Now we see that adding a suitable generic pressure perturbation
to the LTB model can in fact change the outcome of collapse.

>From a physical point of view, the most relevant quadrant
in the model shown in Fig. \ref{Fig}
is the first one (where $p_\theta>0$ and $M_2<0$),
since it corresponds to positive pressures and radially
decreasing energy density profile. We see that both the
black hole and naked singularity outcomes
are possible in this case.

We shall stress however that the model illustrated above,
while presenting a simple and intriguing structure, does
not depict the only possible structure of collapse with tangential
pressures. Other examples, with different choices of the
functions $M$, $b$ and $a$, might provide an entirely
different picture for the final stages of collapse, thus
indicating once more how rich and complex is the description of collapse
resulting from general relativistic analysis.

This model shows also how the black hole formation
process described in the well known homogeneous dust scenario
is not `stable' when arbitrarily small stress perturbations are introduced.
Since, as shown in Fig. \ref{Fig}, it lies
on the critical surface separating the `naked singularity
configuration space' from the `black hole configuration space',
any close neighborhood of the OSD model will contain initial
data leading to both outcomes.

On the other hand, we see that in the context of this model,
inhomogeneous dust collapse
leading either to black hole or naked singularity (thus away from
the critical surface where $\chi_2=0$), is `stable' under small
pressure perturbations, in the sense that it is always possible
to find a small neighborhood of the configuration space
which will develop the same final outcome.

The above considerations illustrate the role pressures can play
towards determining the final fate of collapse. Clearly, depending on
their nature, the non-zero tangential pressures can help create a
naked singularity, or a black hole. It follows that a mere introduction
of pressure in the collapsing cloud, by itself, cannot help us
restore the cosmic censorship. Finally we note that the structure
illustrated for the above model
extends naturally to more general pressure profiles,
still leaving unchanged the general behaviour where both naked singularities
and black holes might generate at the end of collapse.

\section{Genericity of black hole and naked singularity final states}\label{generic}

The consideration above helps us gain an important insight
into the nature and genericity of the black hole and naked singularity
formation in gravitational collapse. Let us take, for example,
a scenario where the term $\chi_2$ is positive, thus leading the collapse
to a naked singularity final fate. In that case, it is clear that
there always exists a small enough perturbation in the initial data,
which is in the form of small changes in the initial density, pressures,
and velocities of the collapsing shells, which will still preserve
by continuity the positive sign for the $\chi_2$ term, thus preserving
the naked singularity final fate of the collapse. A similar
conclusion will hold for the black hole final state of
collapse as well.

In this sense, if a collapse evolution was to develop in a
naked singularity, there is a small neighborhood around that
specific initial set in the configuration space of initial data, which also continues
to take the collapse to a naked singularity. In such a sense, the naked
singularity formation (as well as the black hole formation) is generic
in the gravitational collapse with non-zero tangential pressures case.


In the light of these results, we can see that the idea
that the introduction of pressures
must favor the black hole formation process is not correct, at least
for non-vanishing tangential stresses.
One may argue that this situation arises because we have
neglected radial pressures, a choice that, while still not being
the dust model, is not sufficiently `realistic'.
Nevertheless the same arguments can be extended to the case
of generic pressures (as it will be done in a future work),
to show that collapse scenarios leading to a naked singularity
have a small neighborhood in the configuration space of initial
pressure profiles, that still leads to a naked singularity.

Since the parameters that determine the final fate of
collapse cannot be chosen arbitrarily
if one wishes to obtain a black hole, we can say that a specific
final outcome of collapse requires some tuning in the initial
data in contrast with the possible view that any initial configuration
with pressures will necessary destroy the naked singularity
formation picture and lead to a black hole.

The fact that clearly emerges from the present analysis
is that the scenarios describing the formation of naked singularities
as the endstate of collapse are not mere counterexamples to
cosmic censorship whose relevance is limited and confined.
We can therefore say that the presence of tangential stresses
can `undress' a singularity that was covered in the absence
of stresses, or viceversa, a naked singularity dust model can be
covered by the introduction of a suitable positive pressure.

It is therefore useful to relate the scenario presented
here with the known collapse models without pressures, namely the
dust collapse scenarios. The first collapse model within the framework of general
relativity was studied by Oppenheimer, Snyder and Datt, and it described
the collapse of a massive sphere made of homogeneous dust. The
Lemaitre-Tolman-Bondi models then considered the collapse where inhomogeneous
dust is allowed. It turned out that when inhomogeneities in the density
distributions are allowed, then the appearance of naked singularities takes
place as possible endstate of gravitational collapse. Dust collapse models
are well known and widely studied and the structure of the formation of
trapped surfaces and of the singularity is well established, also
in the case of higher spacetime dimensions
\cite{n-dim-dust}.

The dust collapse is obtained from the above framework,
when we impose $p_r=p_{\theta}=0$. In that case, from equation \eqref{nu}
it follows that $\nu'=0$, which together with the condition $\nu(0)=0$ gives
$\nu=0$. This in turns implies that $G=b(r)$ and $H=\dot{R}^2$.
For inhomogeneous dust in general we have,
\begin{equation}
    \chi_1(0)=-\frac{1}{2}\int^1_0\frac{\frac{M'(0)}{v}+
    b_0'(0)}{\left(\frac{M(0)}{v}+b_0(0)\right)^{\frac{3}{2}}}dv \; .
\end{equation}
We see that inhomogeneous dust can lead to the formation of
naked singularities or black holes, depending on the behaviour of
the mass profile and the velocity profile
\cite{JDM}.
Furthermore, in the marginally bound case it is possible
to evaluate explicitly the apparent horizon from
equation \eqref{t-ah} as
\begin{equation}
    t_{ah}(r)= t_0-\frac{1}{3}\frac{M_1}{M_0^{\frac{3}{2}}}r+
    \left(\frac{1}{4}\frac{M_1^2}{M_0^{\frac{5}{2}}}-\frac{1}{3}
\frac{M_2}{M_0^{\frac{3}{2}}}\right)r^2+o(r^3) \; ,
\end{equation}
while in the bound and unbound cases, the apparent horizon curve
can be read from equation \eqref{t-ah}.

Collapse is said to be homogeneous when $\rho=\rho(t)$. Homogeneity,
together with the initial condition $v(r, t_i)=1$, implies some constraints on the
possible matter models and velocity profiles.
Now let us consider the following three conditions,
\begin{itemize}
  \item[(i)] $M=M_0$,
  \item[(ii)] $v=v(t)$,
  \item[(iii)] $b_0(r)=k$,
\end{itemize}
where the second condition implies a simultaneous collapse,
in which all shells of matter fall in the singularity at the same time.
We can see that homogeneity, dust and the above conditions are
not all mutually independent. In fact, in the case of collapse with
vanishing radial stresses, which is under consideration here,
we can prove the following statements:
\begin{enumerate}
  \item Homogeneity $\Leftrightarrow$ (i) and (ii),
  \item (ii) $\Leftrightarrow$ (i), (iii) and dust.
\end{enumerate}

We can see from this that homogeneous collapse implies
necessarily
dust, while vice versa is not necessarily true (note, however,
that when radial stresses are included it is possible to have
homogeneous collapse also in the case
of a perfect fluid with homogeneous pressures).
Therefore, the Oppenheimer-Snyder-Datt model implies necessarily
that the three conditions above are satisfied.
We note that the simultaneous collapse (which implies homogeneous collapse
in the case when $M=M_0$) implies that all shells fall into the
singularity at the same time $t=t_0$, thus excluding the possibility
of its visibility due to the structure of the apparent horizon in that case.
The quantity $\chi_i(0)$ vanishes identically in this
case for every $i$. In order for
the singularity arising in dust collapse scenarios
to be visible, inhomogeneities in the density distribution must
be allowed for and have to be taken into consideration, thus dropping
the condition (i), which automatically implies the non-occurrence
of a simultaneous collapse.

To see the first statement, it is sufficient to use equation \eqref{rho}
with the requirement of homogeneity. From that it follows that
$v(r,t)=3CM(r)\alpha(t)$,
which together with the requirement that $v(t_i)=1$ implies (i) and (ii).
To see the inverse implication, it is sufficient to put (i) and (ii) into
equation \eqref{rho}. To prove the second statement, we make use of
equation \eqref{F} written in the form,
\begin{equation}
\dot{v}=e^{\nu}\sqrt{\frac{M(r)}{v}+\frac{b(r)e^{2A}-1}{r^2}}=h(t) \; .
\end{equation}
Since $h$ does not depend on $r$, it has the same value
for any $r$, which is also that obtained for $r=0$, namely,
\begin{equation}
h=\sqrt{\frac{M_0}{v}+b_0(0)}\; .
\end{equation}
>From this it follows immediately,
\begin{equation}
    M(r) = M_0, \; g(r,v)=a(r,v)= 0, \; b_0(r)=b_0(0) \; .
\end{equation}
Vice versa, assuming dust, and conditions (i) and (iii) to hold,
we see that $\dot{v}$ does not depend on $r$, which implies
$v(r,t)=\alpha(t)+\beta(r)$.
That together with the requirement $v(t_i)=1$ implies $v=v(t)$.

\section{Gravitational collapse with cosmological constant}\label{lambda}

It is interesting and indeed important to see how the presence of a
non-zero cosmological constant can affect the dynamical evolution of a
gravitationally collapsing massive matter cloud. The interest in models
with cosmological constant arises primarily from the observation of the
accelerating speed in the expansion of the universe, an observed phenomenon
that indicates that the universe might be filled with some repulsive force
known as `dark energy'. One of the possible ways by which dark energy can be
described is through the introduction of a non-zero cosmological term in
Einstein equations.

The earlier works on the subject were constrained by some simplifying
assumptions. Namely the collapsing matter was taken to be a dust cloud or
the function $\nu$ was chosen to have a specific dependence on $t$ and $R$ in
order to simplify the integration of equation \eqref{Gdot} (see \cite{Lambda}).

In the present section we extend the previous results to the most general case
and test how the dynamics and the final stages of collapse of the models
described above are affected by the introduction of the cosmological constant.
In doing so, we consider an example of dynamical collapse in the presence of
tangential stress and cosmological term, that connects directly with the models
described in section \ref{perturb}.

The reason for the interest in the introduction of only tangential stresses in
the collapsing cloud with cosmological constant resides in the fact that, from
equation \eqref{mass}, we see that the matching with an exterior vacuum
background must be done with the well-known Schwarzschild-deSitter
or anti-deSitter spacetimes. On the other hand, the presence of radial
pressures would require the matching with a more
complicated generalized Vaidya metric.


Einstein equations \eqref{p} and \eqref{rho}, in the presence of a
cosmological constant term take the form,
\begin{eqnarray}\label{pL}
\Lambda&=&\frac{\dot{F}}{R^2\dot{R}} \; ,\\ \label{rhoL}
\rho+\Lambda&=&\frac{F'}{R^2R'} \; ,
\end{eqnarray}
while the other equations remain unchanged.
The first one above leads to a Misner-Sharp mass of the form
\begin{equation}\label{FLambda}
    F(r,t)=r^3M(r)+\frac{1}{3}\Lambda R^3 \; ,
\end{equation}
while the energy density remains unchanged.
The energy conditions here imply
\begin{equation}
    \rho\geq 0, \; \rho+\Lambda\geq 0, \; \rho+p_\theta\geq 0 \; ,
\end{equation}
however, all of the above need not be respected now, and we may have
a weaker form of energy conditions holding, depending on what the
sign of the cosmological constant is chosen to be.
The matching with the exterior metric in this case needs to be done with
a spacetime that is asymptotically either Schwarzschild-deSitter or
Schwarzschild-anti-deSitter, depending on
the sign of the $\Lambda$ term.

The integration of Einstein equations proceeds exactly in the
same way as described above, and we therefore obtain,
\begin{equation}\label{potential}
    \dot{v}^2=e^{2\nu}\left(\frac{M}{v}+\frac{1}{3}
\Lambda v^2+\frac{be^{2A}-1}{r^2}\right)=\tilde{V}(r,v) \; ,
\end{equation}
where the quantity $\tilde{V}(r,v)$ can now be interpreted as an
effective potential.
The regions of allowed motion are those for which
$\tilde{V}(r,v)\geq 0$ (since $\dot{v}^2\geq 0$).
The zeros of $\tilde{V}$ correspond to turning points of the dynamics where
the collapse is halted and the shell bounces back. Therefore,
if $\tilde{V}(\bar{r}, v)=0$
for some $\bar{r}$, the cloud will bounce back and not collapse
indefinitely towards the center.
Of course, this situation is not allowed in the dust cases when $\Lambda=0$, $r$
is small and where gravity is the only force acting on the collapsing particles.
Still, bouncing behaviours
were found to be possible in scenarios with tangential stresses
and bound velocity profiles.
In the same manner, whenever $\Lambda\neq 0$, it is possible
to observe such bouncing
behaviours in the dynamics of the collapsing shells.

Typically, such a scenario will become relevant for collapsing clouds
of rather large sizes in the universe, for example, for the
case of a large cluster of galaxies collapsing under its own
gravitational attraction,
or possibly for a large collapsing star cluster. In any case,
for the sizes of a collapsing star
of an average size, the effects of a non-zero
$\Lambda$ may not make any significant impact on the final outcome
of its gravitational collapse.

Following a similar analysis of the time curves as that developed
in section \ref{collapse}, we can evaluate $t(r, v)$ in this case as,
\begin{equation}
    t(r,\bar{v})=\int^1_{\bar{v}}\frac{e^{-\nu}}{\sqrt{\frac{M}{v}+\frac{1}{3}
\Lambda v^2+\frac{be^{2A}-1}{r^2}}}dv \; .
\end{equation}
This leads the numerator of $\chi_1(0)$ (which is the essential
element in determining its sign) to have exactly the same
value as in equation \eqref{chi}:
\begin{equation}\label{chiLambda}
    \chi_1(0)=-\frac{1}{2}\int^1_{0}\frac{\frac{M_1}{v}+
    b_0'(0)+2a_1(v)}{\left(\frac{M_0}{v}+\frac{\Lambda}
{3}v^2+b_0(0)+2a_0(v)\right)^{\frac{3}{2}}}dv \; .
\end{equation}

We can thus see that the presence of the $\Lambda$ term affects
collapse in the sense that it might introduce some positive roots in the
effective potentials, and therefore would give rise to some bouncing behaviours.
But it will not affect the terms that are responsible for the sign of $\chi_1(0)$,
which determines the final outcome of collapse to be a black hole or a naked singularity.
This appears reasonable, when we think of the presence of the cosmological
constant as related to a dark energy in the universe, because the effect
that the dark energy will have on collapse can be that of disrupting and preventing
collapse from happening by causing a bounce. But it cannot influence the
nature of the final fate of a collapse that is uniquely determined by
the initial configurations for densities, pressures and velocities.

Furthermore, the presence of an added positive $\Lambda$-term can lead to
$\frac{M}{v}+\frac{1}{3}\Lambda v^2+\frac{be^{2A}-1}{r^2}>0$ while
$\frac{M}{v}+\frac{be^{2A}-1}{r^2}<0$, thus in fact allowing the evolution for
some configurations that were forbidden in the case of vanishing $\Lambda$.

The visibility of the singularity in the case of collapse with a
cosmological constant differs from the case studied above also in the
structure and formation of the apparent horizon. From equation \eqref{FLambda}
and equation \eqref{horizon} we get
\begin{equation}
    r_{ah}^2\left(\frac{M}{v_{ah}}+\frac{\Lambda}{3}v_{ah}^2\right)=1 \; ,
\end{equation}
from which we see that in general the apparent horizon equation is
a cubic equation in $v$ that can have zero, one or more than one positive
roots, thus affecting considerably the formation of trapped surfaces.
Still it is possible to show
(see \cite{Lambda}),
that the presence of the cosmological constant does not
prevent the possibility of formation of naked singularities, which continues to be
a general feature of gravitational collapse.

It is interesting at this point to analyze some examples of dynamical
behaviours based on the above considerations.
We will thus make a specific, although reasonable, choice of the mass,
pressure and velocity profiles and confront the
effective potential given by equation \eqref{potential} written in the form,
\begin{equation}
    V(r,v)=e^{-2\nu}\dot{v}^2=\left(\frac{M(r)}{v}+\frac{1}{3}\Lambda v^2+\frac{G(r,v)-1}{r^2}\right)
\end{equation}
for different values of $\Lambda$.

Following the examples provided in section \ref{perturb}
we will choose only quadratic terms in the mass and stress profiles
and study a `quasi-Hookean' equation for the pressure of the form
\begin{equation}
    e^{-A(r,v)}=1-\mu_0\frac{r^4}{1+r^2}\left(1-\frac{1}{v^2}\right) \; .
\end{equation}
It is easy to check that in this case $a_0=0$ and
$a_{2n-2}=\sum_{n=2}^{\infty}(-1)^nr^{2n}$ ($n=2,3...$), and we obtain a
model that has the same singular behaviour of that analyzed in equation
\eqref{example}. Therefore, for a suitable choice of the parameters, according
to the inequality given by \eqref{exampleM}, the final outcome of such a scenario
can be either a black hole or a naked singularity. Furthermore, in order to keep the
example as general as possible while maintaining the mathematics behind it
manageable, we will consider bound, unbound and marginally bound collapse
models only in the case of $b_0$ constant.
Finally, for simplicity, in the mass profile we will consider only one
non-vanishing term in the expansion. This choice does not affect the
dynamical behaviour very much, since the lower order terms in the expansion are
the ones responsible for the general features of the mass profile in a
neighborhood of the center.

We therefore choose:
\begin{eqnarray}
  M(r) &=& M_0+M_2r^2 \; ,\\
  b(r) &=& 1+kr^2 \; ,\\
  G(r,v) &=& \frac{1+kr^2}{\left[1-\mu_0\frac{r^4}{1+r^2}
\left(1-\frac{1}{v^2}\right)\right]^2}  \; .
\end{eqnarray}
As seen before the following model can lead to a black hole or a naked
singularity, depending on the values chosen for $M_2$, $\mu_0$ and $k$.
Some plots of the dynamics of this model for a specific choice of the parameters
are given in the figures.
The functions so defined correspond partially to the functions
analyzed by Magli in
\cite{Magli}
(where the cosmological term was not considered) and appear
reasonable in the sense that the pressure is small compared to the energy density.

\begin{figure}[hh]\label{Fig2}
\begin{minipage}[b]{0.5\textwidth}
\centering
\includegraphics[scale=0.4]{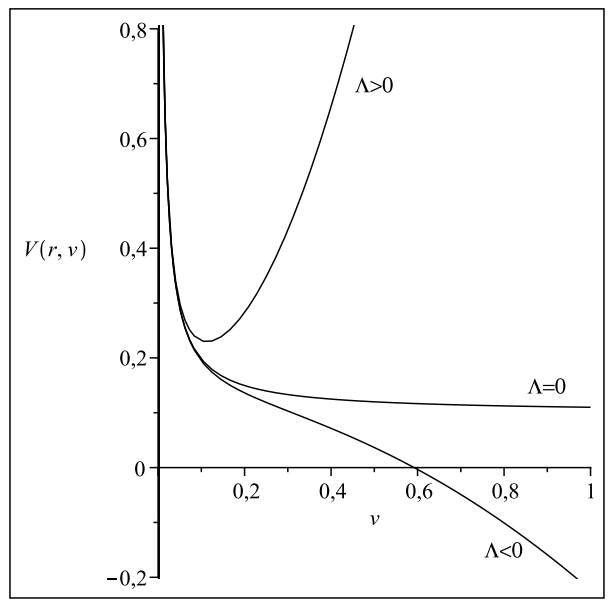}
\end{minipage}
\caption{Sample plot of $V(r,v)$ in the unbound case, at a
fixed value of $r$ close to the center, with $M_0=0.01$, $M_2=-0.001$,
$\mu_0=0.1$ and $k=1$.}
\end{figure}

\begin{figure}[hh]\label{Fig3}
\begin{minipage}[b]{0.5\textwidth}
\centering
\includegraphics[scale=0.4]{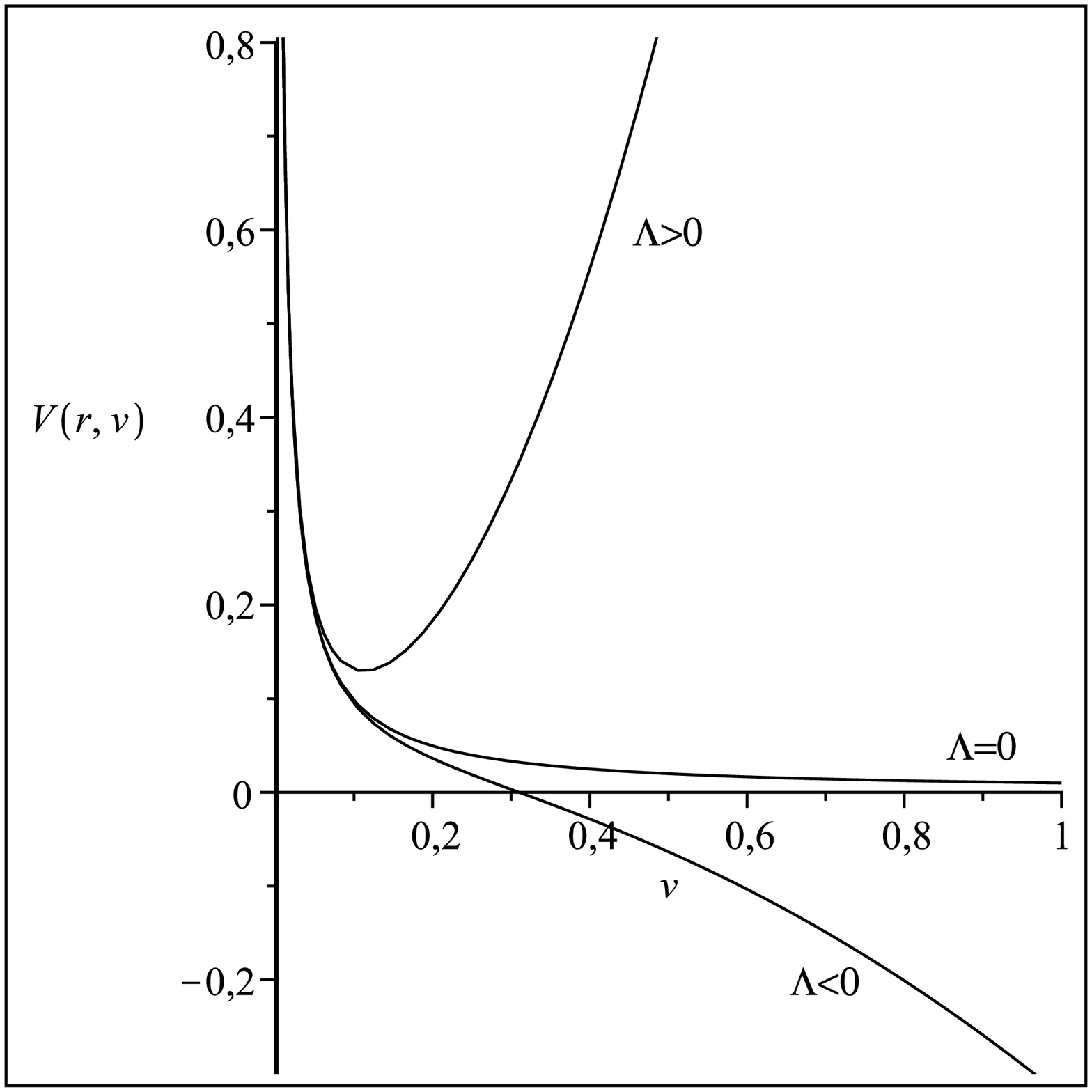}
\end{minipage}
\caption{Sample plot of $V(r,v)$ in the marginally bound case,
at a fixed value of $r$ close to the center, with $M_0=0.01$, $M_2=-0.001$,
$\mu_0=0.1$ and $k=0$.}
\end{figure}

\begin{figure}[hh]\label{Fig4}
\begin{minipage}[b]{0.5\textwidth}
\centering
\includegraphics[scale=0.4]{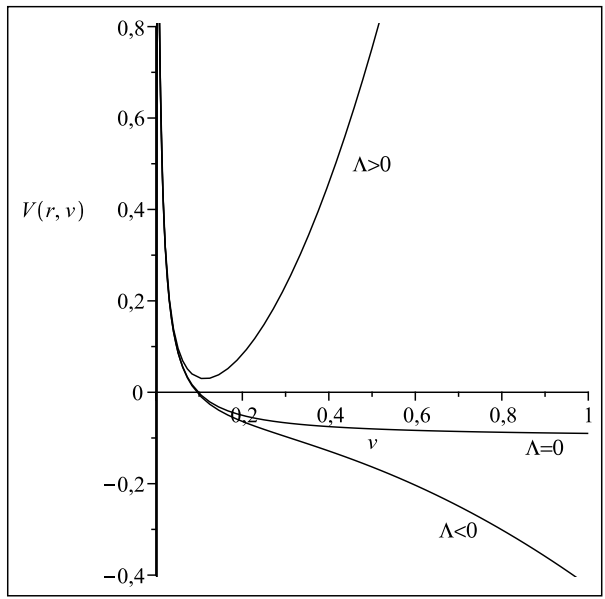}
\end{minipage}
\caption{Sample plot of $V(r,v)$ in the bound case, at
a fixed value of $r$ close to the center, with $M_0=0.01$, $M_2=-0.001$,
$\mu_0=0.1$ and $k=-0.1$.}
\end{figure}

The first thing that can be noted is that the introduction of a
positive cosmological constant can lead to unbound models even in cases
that were bounded in the absence of the $\Lambda$-term, thus making the
dynamics possible for configurations that were previously not allowed. On the contrary,
a negative cosmological constant term might introduce zeroes in $V(r,v)$,
thus altering and limiting the range of allowed configurations.

We can see that, since the dynamics happens in the region of positive
$V$, when $V$ has no zeroes then no bouncing behaviour can happen.
Since in the region close to the center $V$ diverges to plus infinity, we can infer
that the core of the cloud will always collapse to the singularity.
On the other hand, we note that at larger values of $r$ oscillating behaviours
between two fixed radii are possible. Furthermore at larger values of $r$ bouncing
behaviours can happen whenever the shell labeled with $r$ reaches an
event $v$ for which $V(r,v)=0$. In this case the collapse
halts and the dynamics is reversed.
In the figures we have restricted the analysis to a close neighborhood of the
center of the cloud.

As we have noted earlier, the key physical interest for the bouncing behaviour
comes for the collapsing clouds which are large enough in physical scales, but
for stellar collapse, the cosmological constant value plays no significant role.
Of course, even without a cosmological term, a non-zero pressure can cause a
bounce in the outer regions of the cloud.
However, the inner region must still collapse
to the singularity necessarily, where the behaviour of the cloud is very much like the dust
collapse. This is consistent with the regularity condition we mentioned earlier,
namely that the tangential pressure must go to a vanishing value in the limit
of approach to the center. A negative sign for the cosmological term preserves
the same qualitative nature of the collapse as above, however, a positive cosmological
constant brings in interesting changes in the collapse dynamics, as noted above.

\section{concluding remarks}\label{remarks}


The framework to describe collapse of massive objects in general
relativity is extremely rich and complex. Even in the limited case of
spherical symmetry a wide array of different situations
and collapse scenarios arises.

We have here presented a simple and straightforward
analysis of the gravitational collapse of a spherical distribution of mass
in the presence of tangential stresses. We have shown, how the occurrence of
trapped surfaces and the eventual visibility of the singularity
forming at the end of
the collapse process is directly related to the nature and behaviour of
the singularity curve.

This result further supports the conclusion that
for a generic spherical object, the end-state of complete
gravitational collapse will be either a black hole or a naked singularity.
The initial configuration as well as the equations describing
the properties of the matter cloud
during collapse are the only features that, once coupled to
Einstein equations for the dynamics
of the model, will determine the final fate of collapse.

Since the radial and tangential pressures
play different roles during the evolution of collapse
dynamics, the present work
helps understand how much the outcome is
due to either of them. By investigating the
role played by tangential pressures
near the formation of the singularity in full generality, without
any restrictions on the type of matter responsible for
the pressure itself, we allowed
for the possibility to consider negative stresses also.
These, while still satisfying the weak energy condition
(i.e. not considering phantom sources, and therefore constraining
the pressure to a lower finite bound), are possible
candidates for those exotic matter sources that have recently
become of great interest in both cosmology and astrophysics.

In the past years many studies have been devoted to understand the
physical processes involved in the complete gravitational collapse of a massive star.
The treatment presented above offers a comprehensive tool to deal with the
occurrence of naked singularities in the case of vanishing radial stresses and
consequently provides a general picture in which earlier results are included.
As an example, the work by Magli on collapse with tangential stresses
\cite{Magli},
that provides a complete characterization of the outcome of collapse in terms
of the root equation developed by Joshi and Dwivedi, can be translated in the
present framework. As noted, the use of mass-area coordinates
(or the area-radius coordinates) ultimately reduces to the change of variables
$(r,t)\rightarrow(r,v)$, made here in order to study the equation of
motion \eqref{vdot}.
Also, the analysis of the Joshi-dwivedi root equation, which
governs the visibility of the singularity, is equivalent to the analysis
of radial null geodesics exposed in section \ref{collapse}.

Furthermore, the results that give the conditions regarding the
strength of the curvature singularity occurring at the center which were
obtained by Nakao et al. \cite{Nakao} can be translated in the above formalism.
Equations (2.13) and (2.14) in that paper are then found to be equivalent
to equations \eqref{t} and \eqref{vdot} which determine the equation
of motion and the expansion of $t(r,v)$ near the center to first order
in $r$, while equation (2.18) gives a condition for the occurrence of outgoing
null geodesics equivalent to equation \eqref{x0}.

The models provided in section \ref{perturb} and
section \ref{lambda} can be connected
to the ones discussed in the previous papers and therefore
provide a useful tool to broaden the spectrum of collapse
scenarios studied in detail and expand our comprehension
of the final stages of complete gravitational collapse in
general relativity, from the
analytical viewpoint.

The fact that positivity of the first non-vanishing
$\chi_i(0)$ is a necessary and sufficient condition for the
occurrence of naked singularities may be possibly used
in computer simulations, to analyze the endstate of
collapse in a broader and more realistic range of situations.
In fact, the early works in numerical relativity by
Shapiro and Teukolsky \cite{Shapiro} on naked singularity
formation were hindered by the fact that the connection between
the delay of trapped surfaces formation and the existence of
outgoing null geodesics was still unclear. Shapiro and
Teukolsky considered models describing the collapse of prolate spheroids
(thus in connection with the problem of axially symmetric collapse and the
hoop conjecture), and models describing the collapse of counter rotating
particles with vanishing angular momentum (the `Einstein cluster' cited above).
In both cases they found evidence for the non-occurrence of the apparent
horizon before the time of the formation of the singularity.
Still, Wald and Iyer showed
that the fact that on certain slices trapped surfaces form at a later stage does
not necessarily imply the visibility of the central singularity (they pointed out
a time slicing of the Schwarzschild space-time whose evolution approaches
arbitrarily the singularity without the appearance of an apparent horizon
in any slice)
\cite{Wald}.

We can argue here that in comoving coordinates the
occurrence of trapped surfaces and the visibility of the
singularity are related via the sign of $\chi_i(0)$, thus
providing a firm footing for further numerical investigations.
Hopefully, numerical simulations of naked singularity
formation will help shed a better light on the final stages
of realistic collapse and open new possibilities
for the investigations of what happens in last moments of the life of a star.
From an astrophysical point of view, this approach is clearly important,
because if such processes indeed happen in the final stages of
collapse of massive objects, it is then crucial to understand
what kind of signature they bear in order to be able
to eventually `observe' them.


\begin{thebibliography}{99}


\bibitem{PSJ} P.S. Joshi and I.H. Dwivedi, Class. Quantum Grav. \textbf{16},
41 (1999); P.S. Joshi and R. Goswami, Phys. Rev. D \textbf{76}, 084026 (2007).

\bibitem{Penrose} R. Penrose, Riv. Nuovo Cimento \textbf{1}, 252 (1969).

\bibitem{Ref} R. Giamb\'{o}, F. Giannoni, G. Magli and P. Piccione,
Comm. Math. Phys,\textbf{235}, 545 (2003);
M. Celerier and P. Szekeres, Phys.Rev. D \textbf{65}, 123516 (2002);
T. Harada, H. Iguchi and K. Nakao, Prog.Theor.Phys. \textbf{107}, (2002) 449;
P. S. Joshi, Pramana {\bf 55}, 529 (2000);
A. Krolak, Prog. Theor. Phys. Suppl. {\bf 136}, 45 (1999);
R.Goswami, P.S. Joshi, C. Vaz, and L. Witten, Phys. Rev. D \textbf{70}, 084038 (2004).

\bibitem{OSD} J.R Oppenheimer and H.Snyder, Phys. Rev. \textbf{56}, 455 (1939);
S. Datt, Zs. f. Phys. \textbf{108}, 314 (1938).

\bibitem{LTB} G. Lemaitre, Ann. Soc. Sci.Bruxelles I, A \textbf{53}, 51 (1933);
R. C. Tolman, Proc. Natl. Acad. Sci. USA, \textbf{20}, 410 (1934);
H. Bondi, Mon. Not. Astron. Soc., \textbf{107}, 343 (1947).

\bibitem{dust} D. M. Eardley and L. Smarr, Phys. Rev. D \textbf{19}, 2239 (1979);
D. Christodoulou, Commun. Math. Phys., \textbf{93}, 171 (1984);
R. P. A. C. Newman, Class. Quantum Grav., \textbf{3}, 527 (1986);
B. Waugh and K. Lake, Phys. Rev. D \textbf{38}, 1315 (1988);
P. S. Joshi and I. H. Dwivedi, Phys. Rev. D \textbf{47}, 5357 (1993);
S. Jhingan, P. S. Joshi and T. P. Singh, Class.Quant.Grav. \textbf{13}, 3057 (1996).

\bibitem{fluid} A. Ori and T. Piran, Phys. Rev. Lett., \textbf{59}, 2137 (1987);
A. Ori and T. Piran, Phys. Rev. D \textbf{42}, 1068 (1990);
T. Foglizzo and R. Henriksen, Phys. Rev. D \textbf{48}, 4645 (1993);
T. Harada, Phys. Rev. D \textbf{58}, 104015 (1998);
T. Harada and H. Maeda, Phys. Rev. D \textbf{63}, 084022 (2001);
R. Goswami and P. S. Joshi, Class. Quantum Grav. \textbf{19}, 5229 (2002);
P. S. Joshi and R. Goswami, Class. Quantum Grav \textbf{21} 3645 (2004);
R. Giamb\'{o}, F. Giannoni, G. Magli, P. Piccione, Gen. Rel. Grav. \textbf{36}, 1279 (2004);
S. G. Ghosh, D.W. Deshkar, Int. J. Mod. Phys. D \textbf{12}, 913 (2003);
J. F. Villas da Rocha, Anzhong Wang, Class. Quantum Grav. \textbf{17}, 2589 (2000).

\bibitem{press}  A. Mahajan, R. Goswami and P. S. Joshi, Class. Quantum Grav. \textbf{22}, 271 (2005);
P. S. Joshi and R. Goswami, Class. Quantum Grav. \textbf{19}, 5229 (2002).

\bibitem{cross} P. Yodzis, H.-J. Seifert and H. Muller zum Hagen,
Commun. Math. Phys. \textbf{34}, 135 (1973);
C. Hellaby and K. Lake, Astrophysical Journal, \textbf{290}, 381, (1985);
C. J. S. Clarke, The analysis of spacetime singularities, Cambridge University
Press, Cambridge (1993); S. Jhingan and P. S. Joshi, Proceedings of `Internal
Structure of Black Holes and Spacetime Singularities', L.M. Burko and
A. Ori (eds), Annals of the Israel Physical Society, \textbf{13}, (1997) 357.

\bibitem{Magli} G. Magli, Class. Quantum Grav. \textbf{14}, 1937 (1997);
G. Magli, Class. Quantum Grav. \textbf{15} 3215 (1998).


\bibitem{Gonc-Jhin-Mag} S. M. C. V. Goncalves, S. Jhingan, G. Magli, Phys.Rev. D {\bf 65}, 064011 (2002).

\bibitem{Nakao} T. Harada, K. Nakao and H. Iguchi, Class. Quantum Grav \textbf{16}, 2785 (1999).

\bibitem{cluster} S. Jhingan and G. Magli, Phys.Rev. D \textbf{61}, (2000) 124006;
T. Harada, H. Iguchi and K. Nakao, Phys.Rev.D \textbf{58}, (1998) 041502;
H. Kudoh, T. Harada, H. Iguchi, Phys.Rev. D \textbf{62}, (2000) 104016.



\bibitem{Singh} T. P. Singh and L. Witten, Class. Quantum Grav. \textbf{14}, 3489 (1997).



\bibitem{Initial} P.S. Joshi and R. Goswami, Phys. Rev. D \textbf{69}, 064027 (2004).

\bibitem{matching} W. Israel, Nuvo Cemento B \textbf{44}, 1 (1966);
W. Israel, Nuvo Cemento B \textbf{48}, 463 (1966);
A. Wang and Y. Wu, Gen. Relativ. Grav. \textbf{31}, (1), 107 (1999);
P. S. Joshi and I. H. Dwivedi, Class. Quant. Grav. \textbf{16}, 41 (1999);
R Giamb\'o, Class. Quant. Grav. \textbf{22}, 2295 (2005).

\bibitem{Global aspects} P. S. Joshi, \emph{Global aspects in gravitation
and cosmology}, Clarendon Press, OUP, Oxford (1993).

\bibitem{Giambo} R. Giamb\'o, journ. Math. Phys. \textbf{47}, 022501 (2006). 

\bibitem{n-dim-dust} A. Mahajan, R. Goswami and P. S. Joshi, Phys.Rev.D \textbf{72}, 024006 (2005);
R. Goswami and P. S. Joshi, Phys.Rev. D \textbf{69}, 004002 (2004).

\bibitem{JDM} P.S. Joshi, N. Dadhich and R. Maartens, Phys. Rev. D \textbf{65}, 101501(R) (2002). 

\bibitem{Lambda} S. S. Deshingkar, S. Jhingan, A. Chamorro, and P. S. Joshi,
Phys. Rev D \textbf{63}, 124005 (2001); T. Arun Madhav, Rituparno Goswami, and
Pankaj S. Joshi, Phys Rev D \textbf{72}, 084029 (2005).

\bibitem{Shapiro} S.L. Shapiro, S.A. Teukolsky, Phys. Rev. Lett. \textbf{66},
994 (1991); S.L. Shapiro, S.A. Teukolsky, Phys. Rev. D \textbf{45}, 2006 (1992).

\bibitem{Wald} R.M. Wald, V. Iyer, Phys. Rev. D \textbf{44}, 3719 (1991).





\end{thebibliography}
\end{document}